\documentclass[prb,superscriptaddress,twocolumn,showpacs]{revtex4}

\usepackage{graphicx}
\usepackage{dcolumn}
\usepackage{longtable}%
\usepackage{bm} 

\begin{document}

\title{Spin-transfer mechanism of ferromagnetism in polymerized
fullerenes: $Ab \; initio$ calculations}

\date{24 October 2005, Phys. Rev. B \textbf{72}, 214426 (2005) }

\author{O. E. Kvyatkovskii}
\email{kvyatkovskii@mail.ioffe.ru} 
\affiliation{Ioffe Physico-Technical Institute of the RAS, St. Petersburg, 194021, Russia}
\author{I. B. Zakharova} 
\affiliation{State Polytechnic University, St. Petersburg, 195251, Russia}
\author{A. L. Shelankov} 
\affiliation{Ioffe Physico-Technical Institute of the RAS, St. Petersburg, 194021, Russia}
\affiliation{Ume{\aa} University, 90187 Ume{\aa}, Sweden}
\author{T. L. Makarova}
\affiliation{Ioffe Physico-Technical Institute of the RAS, St. Petersburg, 194021, Russia}
\affiliation{Ume{\aa} University, 90187 Ume{\aa}, Sweden}

\begin{abstract}
A mechanism of the high temperature ferromagnetism in polymerized
fullerenes is suggested.  It is assumed that some of the C$_{60}$
molecules in the crystal become magnetically active due to spin and
charge transfer from the paramagnetic impurities (atoms or groups), such
as hydrogen, fluorine, hydroxyl group OH, amino group NH$_2$, or
methyl group CH$_3$, dispersed in the fullerene matrix.  The exchange
interaction between the spins localized on the magnetically active
fullerenes is evaluated using \textit{ ab initio} calculations. The
nearest neighbour and next nearest neighbour exchange interaction is
found to be in the range $0.1\div 0.3 $ eV, that is, high enough to
account for the room temperature ferromagnetism.  
\end{abstract}


\maketitle

\section{Introduction}

Recent developments in material science have resulted in the discovery of a
novel class of magnetic organic materials based on carbon, such as
fullerenes \cite{Allemand91, Ata94, Sato97, Murakami96, Makarova03a,
Owens04, Makarova01, Wood02, Narozhnyi03, Makarova03b} and graphite.
\cite{Esquinazi03,Esquinazi04} Fullerenes C$_{60}$ intercalated with
the organic TDAE-molecule is ferromagnetic below 16 K.
\cite{Allemand91,Sato97} The ferromagnetic ordering below $\approx
$370~K in the PVDF-C$_{60}$ composite, where PVDF is
polyvinylidenefluoride, (-CH$_{2}$-CF$_{2}$-)$_{n}$, has been
reported.  \cite{Ata94} More recently, room temperature ferromagnetism
has been reported in fullerenes polymerized by means of
photo-processes \cite{Murakami96,Makarova03a,Owens04} and by high
pressure -- high temperature treatment
\cite{Makarova01,Wood02,Narozhnyi03,Makarova03b} and
hydrofullerites.\cite{C60H24} In the present paper, we focus our
attention on the problem of the high temperature magnetism in
polymerized fullerenes.

In accordance with the band structure calculations, an ideal lattice
of polymerized fullerenes \cite{Boukhvalov04} is not expected to show
magnetism.  Some yet unidentified structural or chemical imperfections
are of crucial importance for the fullerenes to become magnetic.  The
data suggest that the type of polymerization, rhombohedral
\cite{Makarova01,Wood02,Narozhnyi03} or tetragonal,
\cite{Makarova03b,Spemann03} is not of primary importance, and that
the polymerization is necessary but not sufficient a condition for a
high temperature ferromagnetism.  For photo-polymerized fullerenes,
the presence of oxygen is a
prerequisite. \cite{Murakami96,Makarova03a,Owens04} The results of the
Ref.~\onlinecite{Ata94} give the hint that C$_{60}$ radical adducts,
C$_{60}R_{n}$ where $R$ originate from the organic polymer fragments,
are responsible for the magnetism. Seeing that the intrinsic magnetism
in polymerized fullerenes remains an experimentally controversial 
issue\cite{Spemann03,Hohne02,Han03}, it is of interest to demonstrate
that the high-temperature magnetism is a theoretically feasible
possibility in realistic assumptions about the materials.

To answer the question about a microscopic origin of the
ferromagnetism in carbon-based materials, the theory has to find the
structural element of the carbon matrix that carries unpaired spins,
and to show that the interaction between the spin of the units leads
to a parallel spin alignment.  So far, the theoretical development has
been in the picture of magnetically active structural defects,
\textit{i.e.}, defects with localized spins, interaction of which is
mediated by the magnetically passive carbon
matrix. \cite{Hjort00,Ma04,Lehtinen03,Andriotis03,Kim03,Park03,Chan04}
Defects of different complexity and topology were considered: carbon
vacancies in graphite, \cite{Hjort00,Ma04} carbon adatoms on the
graphene layer, \cite{Lehtinen03} vacancies in the fullerene cages
\cite{Andriotis03},  partially opened intermediate fullerene cage
structures with the zigzag-type edge, \cite{Kim03} the carbon
tetrapods with negative Gaussian curvature \cite{Park03}, and the
special open-cage defect structure with the hydrogen atom bonded
chemically to one of defect carbon atoms. \cite{Chan04} It is yet
unclear whether these defects are present in real fullerene samples
(see, however, Ref. \onlinecite{Hashimoto04}, where atomic-scale
defects were observed in graphene layers) and whether the spin-spin
interaction is ferromagnetic and strong enough to account for the high
temperature magnetism of polymerized fullerenes.

As pointed out in Ref.~\onlinecite{Chan04}, in the periodic version of
the defective structure of Rh-C$_{60 }$ proposed in the
Ref.~\onlinecite{Andriotis03}, \textit{ i.e.}, in Rh-C$_{59}$, the
interaction between the magnetic moments localized on the cages is not
ferromagnetic. The defective structure model proposed in
Ref.~\onlinecite{Chan04} does provide a mechanism for ferromagnetic
ordering, but the computed energy difference between ferromagnetic and
antiferromagnetic states is too small (3 meV per cage) to explain the
high temperature magnetism.  Besides, the magnetically active
structural defects approach is hardly applicable to cold-polymerized
fullerenes \cite{Murakami96,Makarova03a,Owens04,Ata94} where the
formation of C$_{60}$ cage defects is ruled out. A model, where
magnetism arises in a system of undamaged buckyballs, has been
suggested in Ref. \onlinecite{Ribas04}.  In accordance with the
calculations, a neutral C$_{60}$ {\it dimer} turns into the triplet
state provided the interfullerene bond is shortened below the
critical length 1.3 {\AA}. \cite{Ribas04} This value is appreciably
smaller than the experimentally observed bond length 1.58 - 1.62
{\AA}, and an enormous pressure would be needed to decrease the
intermolecular distance below the critical one. It is not clear
whether the model is universal in particular in the view of the
observation that the magnetic transition may take place even without
any external
pressure.\cite{Murakami96,Makarova03a,Owens04,Ata94,Allemand91}

In this paper we suggest a scenario 
where doping creates fullerene radical adducts C$_{60}R$, 
i.e., paramagnetic species with unpaired spins localized
on fullerenes. In our model, ferromagnetism occurs in a network of
the paramagnetic species dispersed in a polymerized fullerene matrix.  
The model is based on the computational observation\cite{KvyMakZac04} 
that the ground state of a doubly charged [2+2] cycloadduct dimer
(C$_{60}^{\pm}$)$^{2}$ is \textit{ triplet}, \cite{KvyMakZac04} with
the singlet state separated by the gap of order of 
$0.2\div 0.3$~eV. \cite{KvyMakZac05} In other words, the spins localized on the
neighboring C$_{60}^{\pm}$ ions interact ferromagnetically, and the
interaction is strong enough to account for the high temperature
ferromagnetic transition in polymerized fullerenes.

In our model we assume that some of the C$_{60}$ molecules in the
crystal, initially spin inert and charge neutral, become magnetically
active (paramagnetic) due to the charge transfer from radicals
(impurities), naturally accompanied by the spin transfer.  Although
the electron transfer is the actual reason for the fullerenes to
acquire spin, we label the scenario ``spin transfer'' because C$_{60}$
acquire spin $\frac{1}{2}$ for any direction -- to or from the
C$_{60}$ molecule -- of the electron transfer. The purpose of this
paper is a detailed study of the spin transfer model using first
principle calculations.

The present mechanism is best illustrated by the case of a
paramagnetic impurity $A$ ($A=$ alkali metals, hydrogen, fluorine, the
hydroxyl group OH, amino group NH$_2$, methyl group CH$_3$ {\it etc} )
allowing for the charge-spin transfer reaction
\begin{eqnarray}\label{eqn1}
\text{paramagnetic }A 
&+& 
\text{diamagnetic C}_{60} 
\hspace*{1ex}\longrightarrow  
\nonumber   \\
\text{diamagnetic }A^{+}(A^{-})  
&+& 
\text{paramagnetic C}_{60}^{-}(\text{C}_{60}^{+})
\hspace*{6ex}
\end{eqnarray}
with the formation of charged paramagnetic ions C$_{60}^{\pm}$. Here
we speak about the final states of the charged impurity or fullerene,
with particular emphasis on their spin states.  One can see that our
scenario has many features in common with the charge transfer
complexes model \cite{Sato97} proposed for the description of magnetic
properties of TDAE-C$_{60}$ and to the qualitative picture of
paramagnetic (open shell) C$_{60}$ radical adducts considered
responsible for the ferromagnetism in C$_{60}$ dispersed in
PVDF. \cite{Ata94}

In the model under consideration, the primary source of the local spin
moments is the radical impurities dispersed in the fullerene lattice.
In this respect, the spin transfer magnetism in polymerized C$_{60}$
is akin to the ferromagnetism in a dilute system of magnetic atoms in
an insulator or semiconductor, with some important differences
however. First, the reaction Eq.~(\ref{eqn1}) transforms paramagnetic
impurity atom (or molecule) to a \textit{diamagnetic} ion with closed
electron shells. This means that the impurities, nominally
paramagnetic, play  only a passive role in our scenario of the sources
of net charges and spins.  Second, both magnetically active centers
(actual paramagnetic species) and magnetically passive ones
(diamagnetic matrix) are constructed from the fullerene molecules.

To validate our physical picture of magnetism, we evaluate the
coupling between paramagnetic species (nearest neighbors and next
nearest neighbors) in the polymerized fullerene matrix.  The role of
the fullerene radical adducts, C$_{60}R$, where $R$=H, F, OH, NH$_2$ or
CH$_3$ in the formation of the ferromagnetic ground state is
studied. For this purpose, the first-principles cluster calculations
of the electronic structure, optimized geometry, and energies of the
two low-lying levels corresponding to the singlet and triplet spin
states for the pairs of ions C$_{60}^{\pm}$ and C$_{60}$ radical
adducts, C$_{60}R$ ($R$=H, F, OH, NH$_2$, CH$_3$) connected by [2+2]
cycloaddition of ``66'' bonds are carried out.  To inspect the range
of the exchange interaction, similar calculations are performed for
the pair of C$_{60}$H radical adducts occupying the opposite (next
nearest) vertexes of the tetragonal tetramer (C$_{60})_{4}$H$_{2}$,
which is a doped fragment of the 2$D$ polymerized tetragonal phase of
C$_{60}$. \cite{Narymbetov03}

The calculations have been carried out in the framework of the density
functional theory (DFT) and \textit{ab initio} Hartree-Fock (HF)
methods.  In the DFT calculations, the hybrid functional of Becke
(B3LYP) \cite{Becke93,Becke88,Lee88} was employed, which includes the
gradient-corrected exchange and correlation functionals along with the
exact exchange. The HF calculations were carried out using the PC
GAMESS version \cite{Granovsky} of the GAMESS (US) QC
package. \cite{Schmidt93} The DFT calculations were carried out using
the PC GAMESS \cite{Granovsky} and the Gaussian 03 suite of
programs. \cite{Frisch03} We exploited the spin-unrestricted method
for both singlet and triplet spin states.  The Gaussian basis sets
employed are 3-21G and 6-31G*. The energy gradient convergence
tolerance was less than 10$^{-4}$ A.U.

Recently, the B3LYP method has been successively applied to solids.
\cite{Muscat01,Bredow00,Xie03,Perry01,Feng04a,Feng04b} A significant
improvement over the LDA results for electronic, structural and
vibrational properties for some semiconductors and insulators, has
been achieved by this method. \cite{Muscat01,Bredow00,Xie03} In
addition, the B3LYP improves the magnetic moments and energy gaps and
correctly predicts the ground state for some antiferromagnetic
insulators.  \cite{Perry01,Feng04a,Feng04b} For the random impurity
distribution case, the cluster approach is more appropriate then
methods developed for periodic systems.

The paper is organized as follows.  In Section \ref{exch}, we consider
a dimer as the smallest nontrivial fragment of the lattice. We study
the charge and spin distributions in the doubly charged dimer
(C$_{60}$)$^{\pm 2}$ and calculate the exchange interaction of the
spins localized on the fullerenes in the dimer. In Section \ref{dop},
we consider spin properties of dimers ``doped'' with radicals.  To
estimate the range of the exchange, we consider in Section \ref{dop} a
cluster with four fullerenes and calculate the next nearest neighbor
exchange interaction.  In the last section, we discuss the results and
estimate the Curie temperature in our model.

\section{Charged dimer: the exchange interaction}\label{exch}

The rhombohedral (Rh) and tetragonal (Tg) polymerized C$_{60}$
crystals are built of C$_{60}$ layers in which fullerene molecules are
connected by [2+2] cycloaddition of ``66''
bonds. \cite{Narymbetov03,Chen02} The smallest lattice fragment is two
adjacent C$_{60}$ molecules, a dimer, and we begin our study with the
analysis of this simplest system in its neutral and charged states.

It is known from the literature that the ground state of a free
fullerene dimer radical adduct $R$--C$_{60}$--C$_{60}$--$R$ is a
singly bonded isomer with the radicals $R$ placed in specific
positions.\cite{Oszlanyi96} The transition from the [2+2] double bond
isomer to the singly bonded one requires breaking of one of the bonds
and a rotation of the buckyball around the remaining bond.  This
process may occur in a metastable dimer phase \cite{Oszlanyi96} but in
a polymeric fullerene network, where the molecules are tightly bound
with the nearest neighbors, the rotation costs the bending and torsion
energy and is, therefore, unfeasible. Thus, we assume that the charge
and spin transfer from impurities to fullerene molecules take place
without changing the bonding type.  Therefore, we study a doubly
charged [2+2] cycloadduct isomer (C$_{60})_{2}^{2\pm}$ as a fragment
of the doped polymerized fullerene lattice.

To obtain a quantitative information on the spin and bonding
configuration, we have performed spin-unrestricted B3LYP (UB3LYP) calculations
of the electronic structure, the total energy, and optimized geometry
for the charged (C$_{60})_{2}^{2\pm}$ dimer in the state with the
total spin $S=0$ (singlet) and $S=1$ (triplet).

To check our methods, we began our calculation with the case of a
neutral dimer. In agreement with earlier results, \cite{Scuseria96} we
found the [2+2]-cycloadduct isomer singlet with $\bm{D}_{2h}$ symmetry
to be the lowest energy state of (C$_{60})_{2}$.  A triplet state of
the doubly bonded neutral dimer lies 2.1 eV higher in energy. In the
singlet state, the doubly charged [2+2] cycloadduct isomer has
$\bm{D}_{2h}$ symmetry. \cite{Scuseria96} The symmetry of the triplet
states has not been studied before.  We have considered the dimer in
triplet state possessing one of the two simplest symmetry elements:
inversion, $\bm{C}_{i}$, symmetry and the mirror, $\bm{C}_{s}$,
symmetry with the plane of symmetry transverse to the dimer axis. The
B3LYP calculations give identical results for the triplet states of
both $\bm{C}_{i}$ and $\bm{C}_{s}$ isomers, and show that the ground
state of [2+2] cycloadduct (C$_{60})_{2}^{2\pm}$ dimer is the
$\bm{D}_{2h}$ isomer in the triplet spin state.

Listed in Tables \ref{tab1} and \ref{tab2}, the results of the
\mbox{B3LYP/6-31G*} calculations for both negatively charged
(C$_{60})_{2}^{2-}$ and positively (C$_{60})_{2}^{2+}$ charged dimers
show that the ground state of a doubly charged [2+2]-cycloadduct
isomer is the triplet spin state of the $\bm{D}_{2h}$ symmetry.  It is
interesting that the properties of the negatively (``electron doped'')
and positively (``hole doped'') charged dimers are rather close, as
seen from the comparison of Tables \ref{tab1} and \ref{tab2}.  Also,
one sees from Tables \ref{tab1}, \ref{tab2} that the results obtained
at 3-21G and 6-31G* levels are in close agreement.

\begin{table}
\caption{\label{tab1} The total energies, lengths and orders of 
the interfullerene bonds for ${\bf D}_{2h}$ isomer (C$_{60})_2^{2-}$ 
in the singlet and triplet states for the optimized structure calculated 
by the use of the UB3LYP hybrid DFT method with the 3-21G and 6-31G* 
basis sets employed}
\begin{ruledtabular}
\begin{tabular}{l|cccc}
basis set & \multicolumn{2}{c}{3-21G} & \multicolumn{2}{c}{6-31G*} 
\\ 
\hline
spin state & singlet & triplet & singlet & triplet 
\\ 
\hline
total energy (A.U.) & -4547.166 & -4547.175 & -4572.430 &  -4572.442
\\ 
bond length (\r{A}) & 1.593 & 1.597 & 1.597 &  1.593
\\ 
bond order          & 0.799 & 0.796 & 0.859   &  0.864 
\\ 
\end{tabular}
\end{ruledtabular}
\end{table}

\begin{table}
\caption{\label{tab2} The total energies, lengths, and orders of 
the interfullerene bonds for ${\bf D}_{2h}$ isomer (C$_{60})_2^{2+}$ 
in the singlet and triplet states for the optimized structure calculated 
by the use of the UB3LYP hybrid DFT method with the 3-21G and 6-31G* basis 
sets employed.}
\begin{ruledtabular}
\begin{tabular}{l|cccc}
basis set & \multicolumn{2}{c}{3-21G} & \multicolumn{2}{c}{6-31G*} 
\\ 
\hline
spin state & singlet & triplet & singlet & triplet 
\\ 
\hline
total energy (A.U.) & -4546.434 & -4546.442 & -4571.765 &  -4571.772
\\ 
bond length (\r{A}) & 1.584 & 1.602 & 1.581 &  1.597
\\ 
bond order          & 0.806 & 0.791 & 0.874   &  0.859 
\\ 
\end{tabular}
\end{ruledtabular}
\end{table}

To identify the nature of the stationary points on the potential
energy surface (a true minimum or a saddle point), we calculated the
vibrational frequencies for the neutral singlet $\bm{D}_{2h}$
(C$_{60})_{2}$ isomer and for both the singlet and triplet ${\bm
D}_{2h}$ (C$_{60})_{2}^{2\pm}$ isomers. The absence of imaginary
vibrational frequencies indicates that all the stationary points are
minima.

A robust feature seen from the \textit{ab initio} dimer calculations
is that the bridging bonds of the [2+2] cycloadduct are not affected
by doping and are almost impenetrable for the spin.  In other words,
each fullerene in a charged dimer possess a well defined spin
localized on the fullerenes.  Hence, one can conveniently describe the
interacting pair in terms of the Heisenberg Hamiltonian,
\cite{Stevens63}

\begin{equation}\label{eqn2}
H = - 2J\bm{S}_{1}\cdot \bm{S}_{2} \; ,
\end{equation}
where $\bm{S}_{i}$ ($\bm{S}^2 =\, \textstyle{\frac34}$)
is the spin vector on the 
$i$-th site of the dimer, and $J$ is the corresponding exchange
integral.
The latter can be presented as  
\begin{equation}
J = (E_{\uparrow \downarrow }-E_{\uparrow \uparrow })/2 \; ,
\label{3ce}
\end{equation}
where $E_{\uparrow \downarrow }$ and $E_{\uparrow \uparrow }$ are the
energies of the dimer in the singlet and triplet states, respectively.

We calculate the exchange integral $J$, from the singlet-triplet
splitting Eq.~\ref{3ce}, having taken the corresponding energies from
Tables \ref{tab1} and \ref{tab2}. The results are listed in Table
\ref{tab3}.  For all the calculation methods we obtained a positive
exchange integral, \textit{ i.e.}, the interaction between the spins
localized on the adjacent fullerenes is ferromagnetic.  Both
spin-unrestricted and spin-restricted B3LYP calculations give the same
value of the singlet state energy and, therefore, the singlet spin
state is not ``antiferromagnetic'' by its nature.  Note that both the
semiempirical AM1 and \textit{ab initio} Hartree-Fock methods, where
correlation effects are ignored, noticeably overestimate $J$.  The
B3LYP hybrid functional calculations, which take the electron
correlations into account along with the exact exchange, are more
accurate and reliable; they give far lower values of the exchange
integral: $J$=0.16 eV for (C$_{60})_{2}^{2-}$ and 0.10 eV for
(C$_{60})_{2}^{2+}$.

This is the key result supporting our model of ferromagnetic
fullerenes: in all the studied cases of the [2+2] cycloadducts
(C$_{60})_{2}^{2\pm}$ , the exchange integral is invariably positive
and the exchange interaction is rather strong.  
We emphasize that the ferromagnetic exchange is an intrinsic property 
of polymerized [2+2] cycloadduct fullerenes.

We note also that the spin intercage interaction is more sensitive to
the relative orientation of the fullerenes then to the intercage
separation: We see from \textit{ ab initio} calculations that for the
relative orientation which corresponds to a \textit{ singly} bonded
dimer, the separation is almost the same as in the [2+2] cycloadduct
case, but the ground state for the singly bonded isomer
(C$_{60})_2^{-2}$ is different: It is spin singlet separated from the
triplet state by the gap $\sim$ 0.7 eV at the B3LYP/3-21G level.

\begin{table}
\caption{\label{tab3}Effective exchange integral 
$J=(E_{\uparrow\downarrow} - E_{\uparrow\uparrow})$ 
for pair of adjacent C$_{60}^{\pm}$ ions in polymerized fullerene.    
The energies $E_{\uparrow\downarrow}$ and $E_{\uparrow\uparrow}$ 
are total energies of low-lying singlet and triplet states for 
corresponding [2+2] cycloadduct charged ${\bm D}_{2h}$ isomers} 
\begin{ruledtabular}
\begin{tabular}{l|cc}
 & \multicolumn{2}{c}{$J$\, (eV) } 
\\ 
\hline
 method & (C\(_{60}\))\(_2^{2-}\)     &   (C\(_{60}\))\(_2^{2+}\)    
\\ 
\hline
AM1         &  0.52 &  0.45  
\\  
B3LYP/3-21G   &  0.12  &    0.10  
\\  
B3LYP/6-31G*   & 0.16
\footnote{The \textit{ab initio} Hartree-Fock/6-31G* 
calculations give in this case the value equal to 0.72 eV} & 0.10  
\\ 
\end{tabular}
\end{ruledtabular}
\end{table}

\section{
Exchange interaction for  radical adducts C$_{60}R$
}\label{dop} 

Most important question is the availability of impurities with the
wanted property to create spins localized on fullerenes in the
polymerized matrix.  Two types of doping can be envisaged, depending
on the impurity character. The first one is a charged complex formed
in the fullerene matrix, comprising a fully ionized impurity bound
primarily by the Madelung electrostatic forces (ionic limit). This is
the case, for example, in TDAE-C$_{60}$, as well as for the alkali
metal doping.  Second, an impurity atom or molecule forms C$_{60}$
radical adduct covalently bound with a carbon atom of the C$_{60}$
cage (covalent limit).  This is the case, for example, for hydrogen,
fluorine, hydroxyl group OH, amino group NH$_2$, or methyl group
CH$_3$.

We will focus our attention on the covalently bounded radicals
(ligands) and especially on hydrogen. Particular interest in hydrogen
doping stems from the fact that hydrogen, a donor, is always present
in fullerene solids in noticeable amounts.  Hydrogen was detected in
rather large concentration (about one hydrogen atom per six fullerene
molecules) in ferromagnetic samples of pressure-polymerized
fullerene. \cite{Chan04} Fluorine is of general interest for it is one
of the very few atoms that has stronger acceptor properties than
C$_{60}$.  Hydroxyl, amino and methyl groups exemplify simplest
molecular radicals.

\begin{figure}
\includegraphics{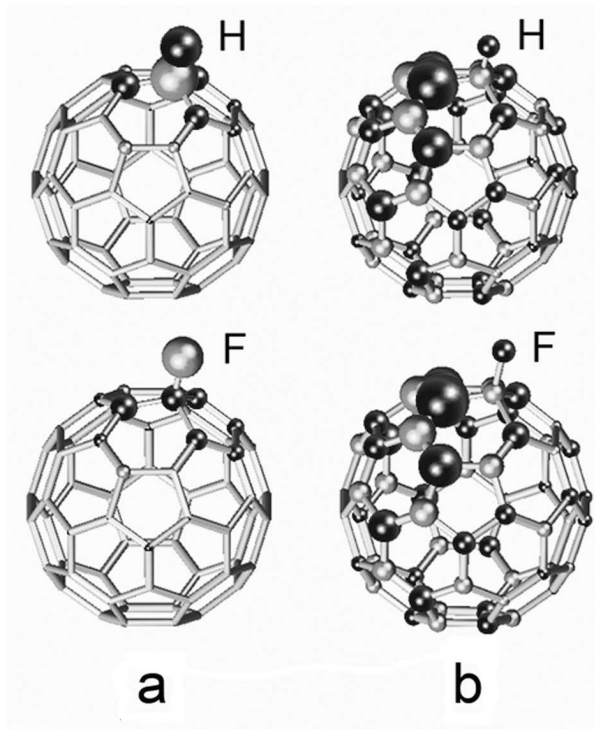}
\caption{
\label{fig1}
The charge (a) and spin (b) density spatial distribution (Mulliken
atomic charges and spins) of the doped fullerenes: C$_{60}$H,
C$_{60}$F. Black (grey) corresponds to positive (negative) charge or up
(down) spins. The volume of the spheres is proportional to the
absolute value of the corresponding variable on a given atom.
}
\end{figure}

We begin with the presentation of our results concerning the
electronic structure the above radical adducts. 
The spatial distribution of the Mulliken atomic charge and atomic 
spin for C$_{60}$H and C$_{60}$F calculated at B3LYP/3-21G level 
is presented in Fig.~\ref{fig1}.
Atomic Mulliken charges (the sum of the atomic Mulliken charges in the
case of fullerene molecule), reflect the nature of the chemical bond,
{\it i.e.}, the degree of hybridization of the valence orbitals of the
fullerene and impurity atoms, rather than the values of atomic net
charges.  The net charges are closely related to the spin density
distribution.  For C$_{60}$H and C$_{60}$F, the B3LYP1/3-21G
calculations show that the Mulliken charges of H and F are equal to
0.27 and -0.25, correspondingly, whereas the atomic spin populations
for both H and F atoms are equal to 0.042.  The
latter values imply that the net charges for H and F are equal to $\pm
0.958$, respectively, and corresponding C$_{60}$ net charges are equal
to $\mp 0.958$.  For comparison, sums of atomic spin populations for
hydroxyl group OH, amino group NH$_2$, and methyl group CH$_3$ are
equal to 0.043, 0.072, and 0.042 respectively, that is much less than
unity as for H and F, whereas corresponding sums of Mulliken atomic
charges are equal to -0.16, -0.03 and 0.15.

\begin{figure}
\includegraphics{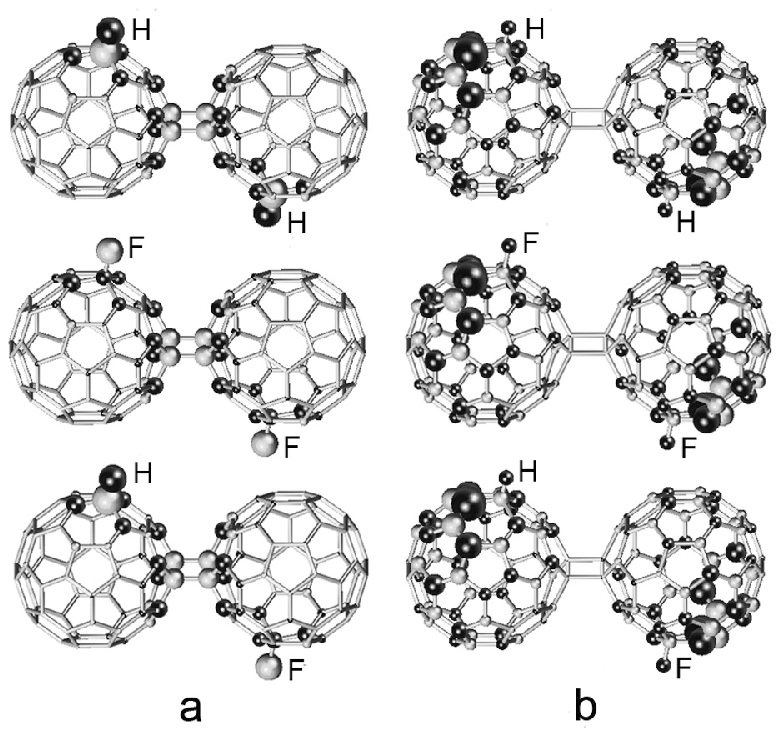}
\caption{\label{fig2}
The charge (a) and spin (b) density spatial distribution (Mulliken
atomic charges and spins) of 
the doped CA dimers (C$_{60}$H)$_{2}$, (C60F)$_{2}$, 
(C$_{60})_{2}$HF.
Black (gray) corresponds to positive (negative) charge or up
(down) spins. The volume of the spheres is proportional to the
absolute value of the corresponding variable on a given atom.
}
\end{figure}

To estimate the exchange interaction between the spins on neighboring
fullerenes in the matrix, we have performed B3LYP/3-21G calculations
for the cases of doped [2+2] cycloadduct dimers
H--C$_{60}$=C$_{60}$--H and F--C$_{60}$=C$_{60}$--F,
OH--C$_{60}$=C$_{60}$--OH, NH$_2$--C$_{60}$=C$_{60}$--NH$_2$,
CH$_3$--C$_{60}$=C$_{60}$--CH$_3$, as well as for a combination of
donor and acceptor H--C$_{60}$=C$_{60}$--F and
H--C$_{60}$=C$_{60}$--OH. Several configurations differing in the
initial position of the ligands relative to the fullerene molecules
have been considered. For all of the configurations, the spins on the
neighboring fullerenes are parallel in the ground state, forming a
triplet, and the energy differences $E_{\uparrow \downarrow } -
E_{\uparrow \uparrow }$ are of the order of several tenths of eV. The
triplet-singlet splittings for the configurations shown in
Fig.~\ref{fig2} are listed in Table~\ref{tab4}.

\begin{table}
\caption{\label{tab4}Total energies for the singlet, $E_{\uparrow\downarrow}$,
and triplet, $E_{\uparrow\uparrow}$, states, and energy gain of 
spin polarized state, $E_{\uparrow\downarrow} - E_{\uparrow\uparrow}$,
calculated at the UB3LYP/3-21G level for a number of doped 
[2+2] cycloadduct $R-$C$_{60}=$C$_{60}-R'$ dimers
and for planar tetragonal tetramer (C$_{60})_4$H$_2$.}  
\begin{ruledtabular}
\begin{tabular}{lccc}
doped isomer& $E_{\uparrow\downarrow}$ & $E_{\uparrow\uparrow}$   & 
$E_{\uparrow\downarrow}-E_{\uparrow\uparrow}$ \\ 
&(A.U.) &(A.U.) & (eV)
\\ 
\hline
H$-$C$_{60}$=C$_{60}-$H              & -4548.182 & -4548.203 & 0.58 
\footnote{The B3LYP/6-31G* calculations give in this case the value equal to 0.57 eV}  
\\ 
F$-$C$_{60}$=C$_{60}-$F              & -4745.555 & -4745.578 & 0.61 
\\ 
H$-$C$_{60}$=C$_{60}-$F              & -4646.869 & -4646.890 & 0.56 
\\ 
OH$-$C$_{60}$=C$_{60}-$OH      & -4697.751 & -4697.773 & 0.59 
\\ 
H$-$C$_{60}$=C$_{60}-$OH          & -4622.967 & -4622.988 & 0.58 
\\ 
NH$_2$--C$_{60}$=C$_{60}$--NH$_2$  & -4658.246 & -4658.268 & 0.58  
\\
CH$_3$--C$_{60}$=C$_{60}$--CH$_3$  & -4626.393 & 4626.414  & 0.57
\\
(C$_{60})_4$H$_2$                    & -9095.234 & -9095.259 & 0.68 
\\
\end{tabular}
\end{ruledtabular}
\end{table}                                                                     

The spatial distributions of the charge and spin density for doped
dimers $R$--C$_{60}$=C$_{60}$--$R'$ (with $R,R'=H,F$) are shown in
Fig.~\ref{fig2}.  The spin density, Fig.~\ref{fig2}(b), is spread
across the buckyballs, repeating the net charge density profile.
Comparing Fig.~\ref{fig2} and Fig.~\ref{fig1}, one concludes that in
the dimer case the spin distribution is nearly the same as for an
isolated C$_{60}R$.  Note that the spin density distribution is
insensitive to the sign of the fullerene charge.  Also, the spin and
net charge distributions are insensitive to the choice of the radical:
Our numerics shows that for $R$--C$_{60}$=C$_{60}$--$R$, $R$= OH,
NH$_2$, CH$_3$, the spatial distributions of the net charge and spin
density over the fullerene molecules are very similar to those in
Fig.~\ref{fig2}.

Remarkably, the spin density is zero on the bonds bridging the
molecules, so that one can assign a spin to each of the buckyballs.
This corresponds to the picture where the spin of an individual
charged (doped) molecule is well defined, and C$_{60}$ molecules play
the role of the sites on which spins reside.  From Table \ref{tab4},
the nearest neighbor exchange interaction Eq.~(\ref{3ce}) is rather
strong, $J \approx 0.3 $~eV.

\begin{figure}
\includegraphics{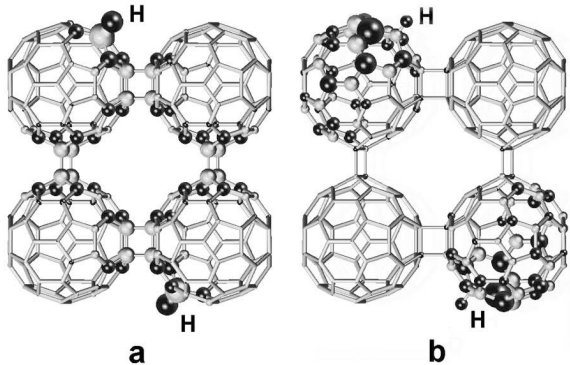}
\caption{\label{fig3}
The charge (a) and spin (b) distribution in the triplet 
spin state of the doped CA tetramer (C$_{60})_{4}$H$_{2}$.
Black (gray) corresponds to positive (negative) charge or up
(down) spins. The volume of the spheres is proportional to the
absolute value of the corresponding variable on a given atom.
}
\end{figure}

To estimate the spatial range of the exchange interaction, we consider
a pair of fullerene hydrogen adducts C$_{60}$H placed in the next
nearest neighbors positions.  For this, we compute properties of
planar tetragonal tetramer (C$_{60})_{4}$H$_{2}$ (Fig.~\ref{fig3})
with doped C$_{60}$ at the opposite corners of the tetramer.  The
UB3LYP/3-21G calculations of the electronic structure and the total
energy have been carried out for optimized geometry of the singlet and
triplet spin states of the cluster. The singlet-triplet splittings are
presented in Table \ref{tab4}. The charge and spin density spatial
distributions for the tetramer (C$_{60})_{4}$H$_{2}$ displayed in
Fig.~\ref{fig3}, are very similar to that for the dimer in
Fig.~\ref{fig2}.  The singlet-triplet splitting is positive
(ferromagnetic) and close to that in a dimer.  Therefore, we observe
that the exchange interaction between fullerene radical adducts in a
2$D$ layer does not fall within the two first coordination
spheres. Moreover, the indirect exchange interaction mediated by
fullerene matrix is somewhat stronger than direct one.  Thus, it is
very likely that the spin interaction extends far beyond the nearest
neighbors, keeping its sign and magnitude.  Of course, additional
calculations of spin interaction between distant fullerene radical
adducts in a large fragments of the lattice are needed to support this
conjecture.

In conclusion of this section, we emphasize that the ferromagnetic
exchange is an intrinsic property of polymerized [2+2] cycloadduct
fullerenes rather than the radical impurity that makes the fullerene
magnetically active.

\section{Discussion and conclusions}

Based on the presented \textit{ab initio} calculations, the following
physical picture emerges. Chemically pure polymerized fullerene
lattice, rhombohedral or tetragonal , is a diamagnet with zero spin on
the sites of the C$_{60}$ matrix.  Doping with radicals initiates
reaction Eq.~(\ref{eqn1}), and some of the cites of the matrix become
paramagnetic as the result of the spin (and charge) transfer from the
radical attached to the corresponding fullerene.  The calculations of
the spin distribution on the doped dimers and tetramers show that the
spin transferred from the radical is smeared over the buckyballs
surface, but it avoids the [2+2] four-membered rings connecting
C$_{60}$'s, so that the transferred spin is well localized on the
fullerenes.  One comes to the picture of immobile spins occupying
sites of the fullerene matrix.

The interaction of the localized spins can be described by the
standard Heisenberg Hamiltonian. In accordance with our calculations,
the Heisenberg exchange integral $J$ is ferromagnetic and rather
strong, $J = 0.1\div 0.3$ eV.  The cluster calculations in Section
\ref{dop}, give the evidence that the ferromagnetic exchange extends
at least to the next nearest neighbor and, probably
further. Polymerized fullerenes, rhombohedral and tetragonal, are
quasi two-dimensional crystals, and it should be noted that the strong
exchange found in our calculations refers only to the in-plane
interaction.  We assume the interplane exchange interaction $J'$ to be
ferromagnetic and weak, similar to that in the low $T_c$
C$_{60}$-TDAE-compounds. Its exact value is not of primary importance, as
discussed below.

In our model, a ferromagnetic transition occurs in a solid solution
$C_{60}R_{x}$ where the impurity (dopants) concentration $x$ is small.
For small $x$, the spatial distribution of the dopant is random so
that the sites of the fullerene matrix become spin active in a random
fashion. Consequently, doped polymerized fullerenes are expected to
belong to the class of disordered ferromagnets. Properties of a
disordered magnet are commonly discussed in the framework of
percolation theory (see Ref. \onlinecite{ShkEfr84} and references
therein).  A detailed analysis is beyond the scope of the paper, and
we limit ourselves to few remarks.

At zero temperature, any two spins are aligned if they are within the
range of the exchange interaction. The long-range magnetic order, when
the aligned spins belong to the infinite cluster, is established
provided the spin concentration, that is the impurity radical
concentration, exceeds the percolation threshold. 
For a two-dimensional square lattice with \textit{only} the nearest neighbors
interaction, the critical concentration $x_{c}$ is $x_{c}\sim
0.59$. If the interaction extends beyond the nearest neighbors, the
critical concentration scales as ${\cal R}^{-2}$ with the radius of
the exchange interaction ${\cal R}$ (measured in units of the lattice
constant).  Our calculations show that the interaction is not limited
to the nearest neighbors, and ${\cal R}$ is not less than $2\sqrt{2}$
and then $x_{c} < 0.07$. In accordance with this estimate, a single
impurity radical for 14, or perhaps more, C$_{60}$ molecules suffices
to transform a diamagnetic polymerized fullerene into a ferromagnet.

The assumption of the percolation theory about a random distribution
of impurities may be invalid for certain synthesis conditions, when
segregation of defects may occur. In this case, the formation of
ferromagnetic islands is expected, with or without bulk magnetic
order.

Evaluation of the Curie temperature, $T_C$, is hindered by the presence
of disorder and a quasi two-dimensional nature of the polymerized
fullerenes.  In accordance with the Mermin-Wagner theorem
\cite{Mermin}, two-dimensional (2D) isotropic magnets exhibit a
long-range order only at $T = 0$ K for any value of the in-plane
exchange integral $J$.  In quasi-two dimensional layered compounds,
the magnetic order establishes at a finite temperature $T_{C}$ due to
the inter-plane exchange $J'$.  It is well-known~\cite{Jongh} that in
the limit of small $J'$, the Curie temperature can be estimated as
$T_{\text C} \sim J/ \ln \frac{J}{J'}$; the estimate should be valid
for our case of a disordered system unless very close to the
percolation threshold.  In accordance with our \textit{ ab initio}
calculations, the in-plane exchange interaction is in the range $J =
0.1\div 0.3 \text{eV} = 1200 \div 3500 $~K .  Seeing that the
dependence of $T_{\text C} $ on $J'$ is only logarithmic, the Curie
temperature $T_{\text C} $ is several times (but not several orders)
less than $J$, \textit{ i.e.}, in the range $300\div 1000 K$.

These estimates, although crude, show that the model has the potential
to give the interpretation to the experiments
\cite{Murakami96,Makarova03a,Owens04,Makarova01,Wood02,Narozhnyi03,Makarova03b}
where magnetism of polymerized fullerenes was observed at room
temperatures and above. These calculation show also that a low
magnetization, which is controlled in our model by the concentration
of the radicals, is compatible with a high Curie temperature.

The presence of any radicals in the studied samples of polymerized
fullerenes is an open question. The only exception is hydrogen, which
was detected in a rather large amount (one hydrogen atom per six
fullerenes) in some polymerized fullerenes exhibiting high-temperature
ferromagnetism.\cite{Chan04}.  Some caution is needed at this point.
Hydrogen is known to have the tendency towards the formation of the
radical adducts C$_{60}H_{n}$ with even number $n$.  Our calculations
show that the ground state of C$_{60}H_{n}$ complex is spin singlet
for $n=2$, and most likely this conclusion holds for higher even
$n$'s.  For this reason, the presence of hydrogen in large amounts
does not guarantee magnetism.  However, if we are concerned with a
\textit{dilute} solid solution of hydrogen, where rare hydrogen atoms
are dispersed chaotically in the fullerene matrix, the formation of
C$_{60}$H complexes becomes feasible during the high-temperature
synthesis.  Formally a metastable state, the system of the C$_{60}$H
complexes quenched upon cooling, is practically stable, being
separated from the true energy minimum by a high energy barrier (of
the order of 2 eV per pair of hydrogen atoms in accordance with our
numerical calculations).  In our scenario, the quenched network of the
magnetically active C$_{60}$H sites becomes ferromagnetic below T$_C$.
The formation of the network is obviously sensitive to details of the
synthesis conditions, and from this point of view, a poor
reproducibility of the magnetic fullerene synthesis procedure can be
understood.

Besides hydrogen, the presence of fluorine, hydroxyl group OH, amino
group NH$_2$, and methyl group CH$_3$, is favorable, in accordance
with our \textit{ab initio} calculation, for the high-temperature
ferromagnetic phase of polymerized fullerenes. Since these materials
have not yet been synthesized and measured, this is a theoretical
possibility waiting for experimental verification.

As has been already mentioned, a similar mechanism of ferromagnetism
is responsible for a low temperature transition of the C$_{60}$
fullerene intercalated with the donor TDAE molecules.  The exchange
interaction is much weaker in this case \cite{Sato97} because the
mutual orientation of the adjacent buckyballs differs from that in the
polymeric phases.

In conclusion, we have considered a mechanism of ferromagnetism in
polymerized fullerenes, where diamagnetic C$_{60}$ molecules transform
into stable paramagnetic species, ions C$_{60}^{\pm }$ or fullerene
radical adducts C$_{60}R$, and become magnetically active due to the
charge and spin transfer from radical impurities. The model is
supported by \textit{ab initio }calculations, the main result of which
is that in the [2+2]-cycloadduct polymerized phase the effective
exchange interaction between the paramagnetic species is ferromagnetic
and strong enough to account for the high temperature ferromagnetism
observed in recent experiments on polymerized fullerenes.  The model
predicts ferromagnetism with the high Curie temperature in the
polymerized fullerenes doped with the radicals like hydrogen,
fluorine, hydroxyl group, amino group, or methyl group.

The work is supported by the Russian Foundation for Basic 
Research (Project 05-02-17779), Swedish Research Council, 
the Royal Swedish Academy of Sciences, and FP6 project 
``Ferrocarbon''.


\begin{thebibliography}{99}

\bibitem{Allemand91} 
  P.-M.~Allemand, K.~C.~Khemani, A.~Koch, F.~Wudl, K.~Holczer, 
  S.~Donovan, G.~Gr\"{u}ner, and J.~D.~Thompson, 
  Science \textbf{253}, 301 (1991).

\bibitem{Ata94} 
  M.~Ata, M.~Machida, H.~Watanabe, and J.~Seto, 
  Jpn.~J.~Appl.~Phys. \textbf{33}, 1865 (1994).

\bibitem{Sato97} 
  T.~Sato, T.~Yamabe, K.~Tanaka, 
  Phys.~Rev. B \textbf{56}, 307 (1997);
  T.~Sato, T.~Saito, T.~Yamabe, K.~Tanaka, and H.~Kobayashi, 
  ibid. \textbf{55}, 11052 (1997).

\bibitem{Murakami96} 
  Y.~Murakami, and H.~Suematsu, 
  Pure~Appl.~Chem. \textbf{68}, 1463 (1996)

\bibitem{Makarova03a} 
  T.~L.~Makarova, K.-H.~Han, P.~Esquinazi, R.~R.~da~Silva, 
  Y.~Kopelevich, I.~B.~Zakharova, B.~Sundquist,  
  Carbon \textbf{41}, 1575 (2003).

\bibitem{Owens04} 
  F.~J.~Owens, Z.~Iqbal, L.~Belova, and K.~V.~Rao,  
  Phys.~Rev.~B \textbf{69}, 033403 (2004).

\bibitem{Makarova01} 
  T.~L.~Makarova, B.~Sundquist, R.~H\"{o}hne, P.~Esquinazi, 
  Y.~Kopelevich, P.~Sharff, V.~A.~Davydov, L.~S.~Kashevarova, 
  and A.~V.~Rakhmanina,  
  Nature \textbf{413}, 718 (2001).

\bibitem{Wood02} 
  R.~A.~Wood, M.~H.~Lewis, M.~R.~Lees, S.~M.~Bennington, 
  M.~G.~Cain, N.~Kitamura, 
  J.~Phys.:~Condens.~Matter \textbf{14}, L385 (2002).

\bibitem{Narozhnyi03} 
  V.~N.~Narozhnyi, K.-H.~M\"{u}ller, D.~Eckert, A.~Teresiaka, 
  L.~Dunsch, V.~A.~Davydov, L.~S.~Kashevarova, and 
  A.~V.~Rakhmanina, 
  Physica B \textbf{329}, 1217 (2003).

\bibitem{Makarova03b} 
  T.~L.~Makarova , B.~Sundqvist, and Y.~Kopelevich, 
  Synth.~Met. \textbf{137}, 1335 (2003).

\bibitem{Esquinazi03} 
  P.~Esquinazi, D.~Spemann, R.~H\"{o}hne, A.~Setzer, K.-H.~Han, 
  T.~Butz,  
  Phys.~Rev.~Lett. \textbf{91}, 227201 (2003).

\bibitem{Esquinazi04} 
  P.~Esquinazi, R.~H\"{o}hne, K.-H.~Han, A.~Setzer, D.~Spemann, 
  T.~Butz,  
  Carbon \textbf{42}, 1213 (2004).

\bibitem{C60H24}
V.E. Antonov, I.O. Bashkin, S.S. Khasanov, A.P. Moravsky,
Yu.G. Morozov, Yu.M. Shulga,
Yu.A. Ossipyan , E.G. Ponyatovskya, 
Journal of Alloys and Compounds \textbf{330-332}, 365 (2002).

\bibitem{Boukhvalov04} 
  D.~W.~Boukhvalov, P.~F.~Karimov, E.~Z.~Kurmaev, T.~Hamilton, 
  A.~Moewes, L.~D.~Finkelstein, M.~I.~Katsnelson, V.~A.~Davydov, 
  A.~V.~Rakhmanina, T.~L.~Makarova, Y.~Kopelevich, 
  S.~Chiuzbaian, and M.~Neumann, 
  Phys.~Rev. B \textbf{69}, 115425 (2004).

\bibitem{Spemann03} 
  D.~Spemann, K.-H.~Han, R.~H\"{o}hne, T.~Makarova, P.~Esquinazi, 
  T.~Butz, 
  Nucl.~Instr.~Meth. B \textbf{210}, 531 (2003).

\bibitem{Hohne02}  
  R.~H\"{o}hne and P.~Esquinazi, 
  Adv.~Mater. \textbf{14}, 753 (2002).

\bibitem{Han03} 
  K.-H.~Han, D.~Spemann, R.~H\"{o}hne, A.~Setzer, T.~Makarova, 
  P.~Esquinazi, T.~Butz.
  Carbon \textbf{41}, 785 (2003).

\bibitem{Hjort00} 
  M.~Hjort, and S.~Stafstr\"{o}m,  
  Phys.~Rev. B \textbf{61}, 14089 (2000).

\bibitem{Ma04} 
  Y.~Ma, P.~O.~Lehtinen, A.~S.~Foster, and R.~M.~Nieminen,  
  New~J.~Phys. \textbf{6}, 1 (2004).

\bibitem{Lehtinen03} 
  P.~O.~Lehtinen, A.~S.~Foster, A.~Ayuela, A.~Krasheninnikov, 
  K.~Nordlund, R.~M.~Nieminen, 
  Phys.~Rev.~Lett. \textbf{91}, 017202 (2003).

\bibitem{Andriotis03} 
  A.~N.~Andriotis, M.~Menon, R.~M.~Sheetz, and L.~Chernozatonskii, 
  Phys.~Rev.~Lett. \textbf{90}, 026801 (2003).

\bibitem{Kim03} 
  Y.-H.~Kim, J.~Choi and K.~J.~Chang, D.~Tomanek, 
  Phys.~Rev.~B \textbf{68}, 125420 (2003).

\bibitem{Park03} 
  N.~Park, M.~Yoon, S.~Berber, J.~Ihm, E.~Osawa, and D.~Tomanek, 
  Phys.~Rev.~Lett. \textbf{91}, 237204 (2003).

\bibitem{Chan04} 
  J.~A.~Chan, B.~Montanari, J.~D.~Gale, S.~M.~Bennington, J.~W.~Taylor, 
  and N.~M.~Harrison, 
  Phys.~Rev.~B \textbf{70}, 041403(R) (2004).

\bibitem{Hashimoto04} 
  A.~Hashimoto, K.~Suenaga, A.~Gloter, K.~Urita, and S.~Iijima,  
  Nature \textbf{430}, 870 (2004).

\bibitem{Ribas04} 
  J.~Ribas-Arino, and J.~J.~Novoa,  
  Angew. Chem. Int. Ed. \textbf{43}, 577 (2004).

\bibitem{KvyMakZac04} 
  O.~E.~Kvyatkovskii, I.~B.~Zacharova,  A.~L.~Shelankov,
  and T.~Makarova, 
\textit{ Proc. of the XVIIIth Int. Winterschool on Electronic
  Properties of Novel Materials},
  AIP Conf. Proc. 723, 385 (2004).

\bibitem{KvyMakZac05} 
  O.~E.~Kvyatkovskii, I.~B.~Zacharova,  A.~L.~Shelankov,
  and T.~Makarova, 
\textit{Proc. of the 7th Int. Workshop on Fullerenes
  and Atomic Clusters, 2005, St. Petersburg, Russia},
  \\
  to be published in Fuller. Nanotub. Car. N..

\bibitem{Narymbetov03} 
  B.~Narymbetov, V.~Agafonov, V.~A.~Davydov, L.~S.~Kashevarova, 
  A.~V.~Rakhmanina, A.~V.~Dzyabchenko, V.~I.~Kulakov, R.~Ceolin, 
  Chem.~Phys.~Lett. \textbf{367}, 157 (2003); 
  M.~N\'{u}\~{n}ez-Regueiro, L.~Marques, J-L.~Hodeau, 
  O.~B\'{e}thoux, and M.~Perroux, 
  Phys.~Rev.~Lett. \textbf{74}, 278 (1995).

\bibitem{Becke93} 
  A.~D.~Becke, 
  J.~Chem.~Phys. \textbf{98}, 5648 (1993).

\bibitem{Becke88} 
  A.~D.~Becke, 
  Phys.~Rev.~A \textbf{38}, 3098 (1988).

\bibitem{Lee88} 
  C.~Lee, W.~Yang, R.~G.~Parr, 
  Phys.~Rev.~B \textbf{37}, 785 (1988).

\bibitem{Granovsky} 
  A.~A.~Granovsky, 
  http://classic.chem.msu.su/gran/gamess/ index.html.

\bibitem{Schmidt93} 
  M.~W.~Schmidt, K.~K.~Baldridge, J.~A.~Boatz, S.~T.~Elbert, 
  M.~S.~Gordon, J.~J.~Jensen, S.~Koseki, N.~Matsunaga, 
  K.~A.~Nguyen, S.~Su, T.~L.~Windus, M.~Dupuis,
  and J.~A.~Montgomery,
  J.~Comput.~Chem. \textbf{14}, 1347 (1993).

\bibitem{Frisch03} 
  M.~J.~Frisch, G.~W.~Trucks, H.~B.~Schlegel, G.~E.~Scuseria, 
  M.~A.~Robb, J.~R.~Cheeseman, J.~A.~Montgomery, Jr., T.~Vreven, 
  K.~N.~Kudin, J.~C.~Burant, J.~M.~Millam, S.~S.~Iyengar, 
  J.~Tomasi, V.~Barone, B.~Mennucci, M.~Cossi, G.~Scalmani, 
  N.~Rega, G.~A.~Petersson, H.~Nakatsuji, M.~Hada, M.~Ehara, 
  K.~Toyota, R.~Fukuda, J.~Hasegawa, M.~Ishida, T.~Nakajima, 
  Y.~Honda, O.~Kitao, H.~Nakai, M.~Klene, X.~Li, J.~E.~Knox, 
  H.~P.~Hratchian, J.~B.~Cross, C.~Adamo, J.~Jaramillo, 
  R.~Gomperts, R.~E.~Stratmann, O.~Yazyev, A.~J.~Austin, 
  R.~Cammi, C.~Pomelli, J.~W.~Ochterski, P.~Y.~Ayala, 
  K.~Morokuma, G.~A.~Voth, P.~Salvador, J.~J.~Dannenberg, 
  V.~G.~Zakrzewski, S.~Dapprich, A.~D.~Daniels, M.~C.~Strain, 
  O.~Farkas, D.~K.~Malick, A.~D.~Rabuck, K.~Raghavachari, 
  J.~B.~Foresman, J.~V.~Ortiz, Q.~Cui, A.~G.~Baboul, S.~Clifford, 
  J.~Cioslowski, B.~B.~Stefanov, G.~Liu, A.~Liashenko, 
  P.~Piskorz, I.~Komaromi, R.~L.~Martin, D.~J.~Fox, T.~Keith, 
  M.~A.~Al-Laham, C.~Y.~Peng, A.~Nanayakkara, M.~Challacombe, 
  P.~M.~W.~Gill, B.~Johnson, W.~Chen, M.~W.~Wong, C.~Gonzalez, 
  and J. A. Pople,
  computer code GAUSSIAN 03, revision B.05, 
  Gaussian, Inc., Pittsburgh PA (2003).

\bibitem{Muscat01} 
  J.~Muscat, A.~Wander, N.~M.~Harrison, 
  Chem. Phys. Lett. \textbf{342}, 397 (2001).

\bibitem{Bredow00} 
  T.~Bredow, and A.~R.~Gerson, 
  Phys.~Rev.~B \textbf{61}, 5194 (2000).

\bibitem{Xie03} 
  R.-H.~Xie, G.~W.~Bryant, and V.~H.~Smith, 
  Phys. Rev. B \textbf{67}, 155404 (2003).

\bibitem{Perry01} 
  J.~K.~Perry, J.~Tahir-Kheli, and William~A.~Goddard III, 
  Phys.~Rev.~B \textbf{63}, 144510 (2001).

\bibitem{Feng04a} 
  X.-B.~Feng, and N.~M.~Harrison, 
  Phys.~Rev.~B \textbf{69}, 035114 (2004).

\bibitem{Feng04b} 
  X.-B.~Feng, and N.~M.~Harrison, 
  Phys.~Rev.~B \textbf{69}, 132502 (2004).

\bibitem{Chen02} 
  X.~Chen, S.~Yamanaka, K.~Sako, Y.~Inoue, and M.~Yasukawa, 
  Chem.~Phys.~Lett. \textbf{356}, 291 (2002).

\bibitem{Oszlanyi96} 
  G.~Oszl\'anyi, G.~Bortel, G.~Faigel, and L.~Gr\'an\'asy, 
  G.~M.~Bendele, P.~W.~Stephens, and L.~Forro,
  Phys.~Rev.~B \textbf{54} 11849 (1996).

\bibitem{Scuseria96} 
  G.~E.~Scuseria, Chem.~Phys.~Lett. \textbf{257}, 583 (1996).

\bibitem{Stevens63} 
  K.~W.~H.~Stevens,  
  in \textit{Magnetism}, edited by G.~T.~Rado and H.~Suhl, Vol. I, 
  (Academic, New York, 1963), p.1.

\bibitem{ShkEfr84} 
  B.~I.~Shklovskii, and A.~L.~Efros \textit{Electronic
Properties of Doped Semiconductors}, Springer Series in Solid-State
Sciences (1984).

\bibitem{Mermin} 
 N.~D.~Mermin and H.~Wagner, Phys.~Rev.~Lett. 
  \textbf{17}, 1133 (1966).

\bibitem{Jongh} \textit{Magnetic Properties of Layered Transition
    Metal Compounds}, edited by L.~J.~de~Jongh (Kluwer, Dordrecht, 1989).


\end{thebibliography}
\end{document}